
\magnification=\magstep1
\hsize 6.0 true in
\vsize 9.0 true in
\voffset=-.5truein
\pretolerance=10000
\baselineskip=13truept

\font\tentworm=cmr10 scaled \magstep2
\font\tentwobf=cmbx10 scaled \magstep2

\font\tenonerm=cmr10 scaled \magstep1
\font\tenonebf=cmbx10 scaled \magstep1

\font\eightrm=cmr8
\font\eightit=cmti8
\font\eightbf=cmbx8
\font\eightsl=cmsl8
\font\sevensy=cmsy7
\font\sevenm=cmmi7

\font\twelverm=cmr12
\font\twelvebf=cmbx12
\def\subsection #1\par{\noindent {\bf #1} \noindent \rm}

\def\mid {\let\rm=\tenonerm \let\bf=\tenonebf \rm \bf}

\def\para{\par \vskip 12 pt}

\def\head{\let\rm=\tentworm \let\bf=\tentwobf \rm \bf}

\def\heading #1 #2\par{\centerline {\head #1} \smallskip
 \centerline {\head #2} \vskip .15 pt \rm}

\def\eight{\let\rm=\eightrm \let\it=\eightit \let\bf=\eightbf
\let\sl=\eightsl \let\sy=\sevensy \let\m=\sevenm \rm}

\def\foots{\noindent \eight \baselineskip=10 true pt \noindent \rm}
\def\sexion{\let\rm=\twelverm \let\bf=\twelvebf \rm \bf}

\def\section #1 #2\par{\vskip 20 pt \noindent {\mid #1} \enspace {\mid #2}
  \para \noindent \rm}

\def\abstract#1\par{\para \foots {\bf Abstract: \enspace}#1 \para}

\def\author#1\par{\centerline {#1} \vskip 0.1 true in \rm}

\def\abstract#1\par{\noindent {\bf Abstract: }#1 \vskip 0.5 true in \rm}

\def\sqr#1#2{{\vcenter{\vbox{\hrule height.#2pt
  \hbox {\vrule width.#2pt height#1pt \kern#1pt
  \vrule width.#2pt}
  \hrule height.#2pt}}}}

\def\n{\noindent}
\def\s{\smallskip}
\def\m{\medskip}
\def\b{\bigskip}
\def\c{\centerline}

\def\gne #1 #2{\ \vphantom{S}^{\raise-0.5pt\hbox{$\scriptstyle #1$}}_
{\raise0.5pt \hbox{$\scriptstyle #2$}}}

\def\ooo #1 #2{\vphantom{S}^{\raise-0.5pt\hbox{$\scriptstyle #1$}}_
{\raise0.5pt \hbox{$\scriptstyle #2$}}}


\voffset=-.5truein
\vsize=9truein
\baselineskip=21pt
\hsize=6.0truein
\pageno=1
\pretolerance=10000
\def\n{\noindent}
\def\s{\smallskip}
\def\b{\bigskip}
\def\m{\medskip}
\def\c{\centerline}

\line{\hfill IUCAA - 17/95}

\line{\hfill June 1995}
\m
\m
\c{\bf\mid STRING-DUST DISTRIBUTIONS WITH }
\c{\bf\mid THE KERR-NUT SYMMETRY}
\b
\b
\n L.K. Patel$^{1} $, {\bf N. Dadhich}$^{2} $ and A. Beesham$^1 $.
\s
\s
\item{1.} Department of Applied Mathematics, University of Zululand,
Private Bag X1001, Kwa-Dlangezwa 3886, South Africa.
\b
\item{2.} Inter-University Centre for Astronomy
\& Astrophysics, Post Bag 4,
 Ganeshkhind, Pune - 411 007, India.
\b
\b
\b
\c{\bf Abstract}
\b
We attempt to solve the Einstein equations for string dust and null flowing
radiation for the general axially symmetric metric, which we believe is
being done for the first time. We obtain the string-dust and radiating
generalizations of the Kerr and the NUT solutions. There also occurs an
interesting case of radiating string-dust which arises from string-dust
generalization of Vaidya's solution of a radiating star.
\b
\b
\b
\b
\b
\b
\b
\n Key words : String dust, Radiating Kerr and NUT, Null Radiation,
Radiating String dust.
\b
\b
\n PACS No : 04.20Jb, 0.4.20 Cv,
\vfill\eject

\item{\bf 1.} {\bf Introduction}
\s
Two empty space solutions of Einstein's equations admitting twisting,
shear-free and geodetic null rays are well-known in the literature.
They are the Kerr solution [1] and the so called Taub-NUT solution
[2]. They are remarkable solutions, the former has important
astrophysical applications while the latter is interesting on the
formal grounds. Vaidya [3] has expressed the usual Kerr metric in
the form

$$\eqalign{ ds^2 &= 2(du + k sin^2 \alpha d \beta) dt - (r^2 + k^2
cos^2 \alpha) (d \alpha^2 + sin^2 \alpha d \beta^2) \cr
& - \bigg(1 + {2mr \over r^2  + k^2 cos^2 \alpha} \bigg) (du + k
sin^2 \alpha d \beta)^2 \cr} \eqno (1.1) $$

\n where $u = t-r $ and $m $ and $k $ are arbitrary constants indicating
mass and rotation parameters of the Kerr particle. On the other hand
the NUT solution is given by

$$\eqalign{ ds^2 &= 2(du - 2b cos \alpha d \beta) dr - (r^2 + b^2)
(d \alpha^2 + sin^2 \alpha d \beta^2) \cr
&- \bigg(1 + {2 mr - 2b^2 \over r^2 + b^2 } \bigg)  (du - 2b cos
\alpha d \beta)^2 \cr} \eqno (1.2) $$

\n where constants $m $ and $b$ are the parameters of the NUT source.
It can be easily verified that the Schwarzschild solution will result
when $k = 0 $ in (1.1) or $b = 0 $ in (1.2). The Kerr solution is
asymptotically flat while the NUT is not.
\s
Further Vaidya et al [4] have synthesized the two into one by writing
$$ ds^2 = 2(du + g ~sin \alpha d \beta) dx - M^2 (d \alpha^2 +
sin^2 \alpha d \beta^2) - 2L (du + g ~sin \alpha d \beta)^2 \eqno (1.3) $$

\n with $g = g (\alpha) $, $M $ and $L $ are functions of $u, x $ and
$\alpha $. The coordinate $x = t $ for the Kerr and $x = r $ for the NUT,
and $u = t - r $ always. In this framework they have obtained some
radiating solutions; i.e. Kerr or NUT source with outflowing null radiation.
\s
In this paper we wish to find for the metric (1.3) solutions of the equation

$$ R_{ik} = -8 \pi K (u_i u_k - g_{ik} - w_i w_k) - 8 \pi \sigma
\zeta_i \zeta_k \eqno (1.4) $$

\n where $u_i u^i = 1 = -w_i w^i $ and $\zeta_i \zeta^i = 0 = u_i w^i $.
The second term represents energy-momentum of the flowing null radiation of
density $\sigma$ while the first corresponds to the energy-momentum tensor

$$ T_{ik} = K (u_i u_k - w_i w_k) \eqno (1.5) $$

\n which is supposed to describe a string-dust distribution [5]. The
string-dust generalization of the Schwarzschild field is well known and
some exact solutions  have been obtained [6-8] of the equation (1.4).
It is perhaps for the first time axially symmetric (metric (1.4))
solutions of this equation are being attempted.
\s
Alternatively one of us [9] would like to interpret the first term in (1.4)
as defining the Machian vacuum. Eqn.(1.5) implies $R_{ik} u^i u^k = 0 $
and $ T^0_0 = T^1_1 = K $, the rest of $T^k_i = 0 $. This means that
the gravitational charge $(\rho + 3p) $ density of the distribution (1.5)
is zero implying no influence on free particles. Though it does not
produce gravitational force on free particles, the distribution (1.5)
does produce non-zero curvature which in spherical symmetry corresponds
to constant (non-zero) gravitational potential [9,10]. It is argued that
 by defining the vacuum by the first term in (1.4), the Schwarzschild
field can be made consistent with Mach's Principle in the sense that
homogeneous and isotropic matter distribution in the Universe lying
exterior to a spherical cavity centred at the mass point can manifest
itself by producing a constant potential. This is why it is called the
 equation for the Machian vacuum [9].
\s
The consideration of eqn. (1.4) can also be motivated by the fact that
very close to the big-bang singularity, the Universe is in highly dense
state and hence its matter content can have very unusual and exotic
properties to allow for viscous effects, heat flux, null radiation flow,
string-dust etc. Some cylindrical models incorporating such dissipative
effects have been considered [11-14]. Here we wish to find the axially
symmetric solutions of eqn. (1.4).
\b
\item{\bf 2.} {\bf Field Equations}
\s
For the metric (1.3) we introduce the orthogonal tetrads as

$$ \theta^1 = du + g sin \alpha d \beta, ~ \theta^2 = M d \alpha, $$

$$ \theta^3 = M sin \alpha d \beta, ~ \theta^4 = dx - L \theta^1 $$

\n and in what follows all the quantitites will be referred to the
tetrad frame.  The Ricci components of the metric (1.3) are given in
the Appendix I [4]. Let us note that $g_{\alpha} = \partial g/\partial
x^{\alpha}, ~ M_{xu} = \partial^2 2M/\partial x \partial u $ etc.
and

$$ 2f = g_{\alpha} + g cot_{\alpha} \eqno (2.1) $$

\n and a new variable $y $ is defined by $g d \alpha = dy $.
\s
We employ the comoving coordinates to write

$$ u_a = (1, 0, 0, 1/2),~ w_a = (1,0, 0, -1/2),~ \zeta_a = (1, 0, 0, 0) $$

\n where the null radiation is taken to flow along the $\theta^1 $ - direction.
\s
For the metric (1.3), eqns. (1.4) imply $R_{23} = 0 $ and $R_{22}
= R_{33} $ and the following system of equations.

$$ R_{44} = 0,~~ R_{42} = 0, ~~ R_{43} = 0 \eqno (2.2 ) $$

$$ R_{41} = 0 \eqno (2.3) $$

$$ R_{12} = 0, ~~ R_{13} = 0 \eqno (2.4) $$

$$ R_{22} = - 8 \pi K \eqno (2.5) $$

$$ R_{11} = - 8 \pi \sigma \eqno (2.6) $$

\n Eqns. (2.2) involve the only one function $M $ (App. I) and their
solution is given by

$$ M^2 = {f \over Y} (X^2 + Y^2) \eqno (2.7) $$

\n where $X = X(x, u, y) $ and $Y = Y(u, y) $ and they satisfy the
conditions

$$ X_x = -1, ~~X_y = Y_u,~~ X_u = -Y_y. \eqno (2.8) $$

Next consider (2.3) which can be solved for the metric function $2L $,

$$ 2L = -{Y_u \over Y} X + 2G + {2FX + 2 EY \over X^2 + Y^2} \eqno (2.9) $$

\n where $E, F, $ and $G $ are functions of $u $ and $y $ satisfying the
relation

$$E = -2YG - YY_y \eqno (2.10) $$

\n From (2.7) - (2.10), it follows after a lengthy algebraic manipulations,

$$ E_u = F_y, ~~ E_y = -F_u \eqno (2.11) $$

\n Then the string density is given by (2.5),

$$ 8 \pi K = {1 \over (X^2 + Y^2)} \bigg[ 2G + {Y \over f}
\lbrace {g^2 \over 2} \bigtriangledown^2 ln (Y/f) - f_y + 3f {Y_y \over Y} + 1
\rbrace \bigg] \eqno (2.12) $$

\n where $\bigtriangledown^2 = \partial^2/\partial u^2 +
\partial^2/\partial y^2 $. Using the above relations, (2.6) will
give the null radiation density $\sigma $, the expression for which
is quite lengthy and it is given in the Appendix II.
\s
For finding the explicit solutions we have to obtain $X, Y $ and $f $
from eqns. (2.8), (2.10) and (2.11). For further investigation we shall
assume $f = Y $.
\b
\item{\bf 3.} {\bf The case $f = Y $}
\s
If $f = Y $, then $Y $ becomes a function of $y $ alone. Eqns. (2.7)
and (2.8) then give

$$ M^2 = X^2 + Y^2, ~~X = au - x, ~~Y = -ay + b \eqno (3.1) $$

\n where $a $ and $b $ are constants of integration.
\s
\n Eqns. (2.8), (2.10) and (2.11) will then lead to

$$ Y \bigtriangledown^2 G - 2 a G_y = 0 \eqno (3.2) $$

\n of which we take the particular solution,

$$ 2G = const. = c \eqno (3.3) $$

\n Then (2.11) will give

$$ E = (a-c) Y, ~ F= a(a-c) U + N \eqno (3.4) $$

\n where $N $ is a constant of integration.
\s
Finally we need to solve the equation implied by the assumption
$f = Y $, which in view of (2.1) will read as

$$ (1 - z^2) Y_{zz} - 2 z Y_z + 2aY = 0,~~ z = cos \alpha \eqno (3.5) $$

\n As seen by Vaidya et al [4], this equation admits a power series
solution for $a \geq - 1/8 $. Writing $2 a = n(n +1) $, it takes to
the familiar Legendre equation,

$$ (1-z^2) Y_{zz} - 2z Y_z + n(n + 1) Y = 0 \eqno (3.6) $$

\n which can be solved for $Y $. In view of $g d \alpha = dy, $ (3.1)
gives the remaining metric potential,

$$ g = -{1 \over a} Y_{\alpha} = {1 \over a} Y_z sin_{\alpha} \eqno (3.7) $$

\n Thus the metric (1.3) is completely determined for the field
equations (1.4). The string dust and the null radiation densities
are then given by

$$ 8 \pi K = - \bigg({2a - c - 1 \over X^2 + Y^2 }\bigg) \eqno (3.8) $$

\n and

$$ 8 \pi \sigma = {2a (a-c) \over X^2 + Y^2} \eqno (3.9) $$

It should be noted that $K = 0 $ for $1 + c = 2a $ and we recover the
radiating case of Vaidya et al [4]. On the other hand $\sigma = 0 $
for $a = 0 $ or $a = c $ presenting a string-dust spacetime.
\vfill\eject
\item{\bf 4.} {\bf The Kerr-like solutions}
\s
When $n $ is a positive integer, we can take the solution of (3.6)
in the form

$$ Y = A P_n (z) + B Q_n (z) \eqno (4.1) $$

\n where $A $ and $B $ are arbitrary constants, and $P_n (z) $ and
$Q_n(z) $ are respectively the Legendre  and associated Legendre
polynomials of order $n $.
\s
\n The metric (1.3) is then given by

$$\left.
\eqalign{& 2 L = c + {2(a-c) (Y^2 + auX) + 2NX \over X^2 + Y^2} \cr
& g sin \alpha = {1 \over a} \bigg(A {d P_n \over dz} + {B d Q_n
\over dz} \bigg) sin^2 \alpha  \cr} \right\} \eqno (4.2) $$

\n where $X = au - x $ and $Y $ as given in (4.1). Here $x $ is a
timelike coordinate and hence replace $x $ by $t $ in $X $ as well
as in (1.3).
\s
When $1 + c = 2a $, the string-dust density vanishes and we recover
the radiating Kerr metric [4] for $B = 0 $ while $A = 0 $ gives the
associated radiating Kerr metric [15]. The associated Legendre function
$Q_n (z) $ has a singularity on the axis $\alpha = 0 $, so would have
the associated metric. The radiation density vanishes for $a = 0 $ or
$a = c $ but $a = 0 $ leads to $g = 0 $ in view of (3.1) and (2.1) and
hence $a = c $ for the radiation free string-dust solution. Then

$$ 8 \pi K = {1 - c \over X^2 + Y^2} \eqno (4.3) $$

\n which means $c \leq 1 $ for $K \geq 0 $. Here we have $X = c u - t ,
{}~~ Y = -cy + b $, the time dependence in $X $ is spurious can be removed
 by $X = R = (c-1) t- cr $, the new radial coordinate. We now take $B = 0 $
to seek the string-dust generalization of the Kerr metric, which results
when $c = 1, $ i.e. $n = 1 $ and from (4.1) $Y = A P_1 (z) = A cos \alpha $.
 For the string-dust we shall hence have to take $n > 1 $, say $n = 2 $
then $Y = A P_2 (z) = {A \over 2} (3 cos^2 \alpha - 1) $. Note that it
would not be possible straight way to get to the Kerr metric from the
string-dust because (4.1) admits different solutions in the two cases for
$n = 1, 2 $. Since $n = 2 $ implies $c = 3 $, from (4.3) the string density
$K $ will be negative and so would be the case for all $n > 1 $.
Following the standard procedure [16], we can bring the metric to
the standard Boyer-Lindquist form,

$$\eqalign{ ds^2 &= {2c - \lambda \over c^2} dt^2 - (R^2 + Y^2)
(9A^2 cos^2 \alpha sin^2 \alpha + (R^2 + Y^2) (2c - \lambda)
)^{-1} d R^2 \cr
&- (R^2 + Y^2) d \alpha^2 - (R^2 + Y^2 + 9 {A^2 \over c^2}
\lambda cos^2 \alpha sin^2 \alpha) sin^2 \alpha d \beta^2 \cr
&+ 6 {A \over c^2} (c - \lambda) cos \alpha sin^2 \alpha dt d
\beta \cr} \eqno (4.4) $$

\n where

$$ \lambda = c + {2mR \over R^2 + Y^2} \eqno (4.5) $$
\s
Similarly one can easily obtain the metric for the string-dust
generalization of the associated Kerr metric. Clearly $A $ is the
rotation parameter and $m $ is the mass. $A = 0 $ will give the
Schwarzschild string-dust. The above metric apparently has many
interesting properties, such as the inherent angular velocity $w =
-g_{03}/g_{33} $ vanishes at both $\alpha = 0 $ and $\alpha = \pi/2 $,
 which will be considered separately [17].
\b
\item{\bf 5.} {\bf The NUT- like solutions.}
\s
We now consider the equation (3.6) for $0 ~\leq ~n \leq 1 $, a particular
 solution of which is given by [4],

$$ Y = b [1 - n (n +1) p_n (z) ] \eqno (5.1) $$

\n where $p_n (z) $ stands for the sum of the infinite convergent series,

$$ {1 \over 2} z^2 + {1 \over 4!} (2 - n) (3 + n) z^4 + {1 \over 6!}
(2-n) (4-n) (3 + n) (5+ n) z^6 + \ldots \eqno (5.2) $$

\n The metric functions $2L $ as given by (4.2)  and

$$ g sin \alpha = -2b {d p_n \over dz} sin^2 \alpha \eqno (5.3) $$

\n where $X = au - r $ (replacing the spacelike coordinate $x $ by $r $)
and $Y $ as given by (5.1). The solution is hence given by $2L $ in (4.2),
(5.1) and (5.3), with $2a = n(n + 1) $. The series (5.2) diverges as $z
\longrightarrow 1 $ and consequently the solution has a singularity
at $ \alpha = 0 $, for all $n \not= 0 $. The axis is thus singular.
The densities are as given by (3.8) and (3.9) with $X $ and $Y $ as
given above. This is the string-dust generalization of the radiaing
NUT metric, discussed by Vaidya et al [4].
\s
When $a = 0 $, i.e. $n = 0 $, the radiation density $\sigma = 0 $
(recall in the Kerr case the appropriate condition for $\sigma = 0 $
was $ a = c $, while here it is $a = 0 $) and in that case the metric
will describe the NUT with string-dust and is given by

$$
\eqalign{ds^2 &= 2(du - 2b cos \alpha d \beta) dr - (r^2 + b^2)
(d \alpha^2 + sin^2 \alpha d \beta^2) \cr
&- \bigg(c + 2 {mr - cb^2 \over r^2 + b^2} \bigg) (du - 2b cos
\alpha d \beta)^2 \cr} \eqno (5.4)
 $$

\n where $N = m $. The string density is now given by

$$ 8 \pi K =  {c + 1 \over r^2 + b^2} \eqno (5.5) $$

\n The above metric can be transformed to the $BL $ form to read as

$$\eqalign{ ds^2 &= \lambda dt^2 - \lambda^{-1} dr^2 - (r^2 + b^2)
d \alpha^2 - (r^2 + b^2 + 4 \lambda b^2 cos^2 \alpha sin^2 \alpha ) \cr
& sin^2 \alpha d \beta^2 + 4 \lambda b cos \alpha sin^2 \alpha dt d
\beta \cr} \eqno (5.6) $$

\n where $\lambda = {- cr^2 + 2mr + cb^2 \over r^2 + b^2 },~ -1 \leq c$ and
hence $K \geq 0 $. When $c = -1 $, it reduces to the
usual NUT metric on the other hand $b = 0 $ implies vanishing of
rotation of the null congruence and we get the Schwarzschild string-dust.
\b
\item{\bf 6.}{\bf Discussion}
\s
There is an important difference between the string-dust generalizations
of the Kerr (4.4) and the NUT (5.6). From (4.4), the Kerr metric does not
result when string-dust is switched off by putting $c = 1 $, while (5.6)
yields the usual NUT metric for $c = -1 $. It would not be possible to
match (4.4) continuously to the Kerr metric where as (5.6) will match
continuously across the boundary, $r = r_0 $ to the NUT metric with
mass parameter $\overline m $ being given as

$$ \overline m = m + {1 + c \over 2r^2_0} (r^2_0 + b^2) \eqno (6.1) $$

We shall now consider an important particular case. Consider $g = 0 $,
the spacetime (1.3) then admits a null congruence which is geodetic as
well as shear and twist free. Further take $M = M(r), ~ L = L(r, u),~
 u = t-r $, then $R_{44} = 0 $ determines $M = r $ and eqns. $R_{24}
= R_{34} = R_{12} = R_{13} = 0 $ become identities. Eqn. $R_{14} = 0 $
integrates to give

$$ 2L = c(u) - 2 {m (u) \over r} \eqno (6.2) $$

\n where $c (u) $ and $m (u) $ are arbitrary functions of $u $. We have

$$ 8 \pi K = {c-1 \over r^2},~~ 8 \pi \sigma = {2m_u \over r^2} -
{c_u \over r} \eqno (6.3) $$

\n and the metric reads

$$ ds^2 = 2 du dt - \bigg(c (u) - {2 m(u) \over r} \bigg) du^2 -
r^2 (d \alpha^2 + sin^2 \alpha d \beta^2). \eqno (6.4) $$

\n This is the string-dust generalization of Vaidya's radiating star
solution, which results when $c = 1 $. The string-density will not
depend upon $t $ for $c = const. \not= 1 $. The Schwarzschild
string-dust will follow when $m $ and $c $ are constants. The
interesting case occurs when $m = const. $ but $c = c (u) $. This
is the radiating string-dust.
\b
\n {\bf Acknowledgement :} One of the authors (LKP) would like to thank
 the University of Zululand, South Africa for hospitality.

\vfill\eject
\n{\bf References}
\s
\item{[1]} R.P. Kerr, Phys. Rev. Lett. {\bf 11} (1963) 237.
\s
\item{[2]} E. Newman, L. Tamburino and T. Unti, J. Math. Phys. {\bf 4}
 (1963) 915.
\s
\item{[3]} P.C. Vaidya, Proc. Cambridge Phi. Soc. {\bf 75} (1974) 383.
\s
\item{[4]} P.C. Vaidya, L.K. Patel and P.V. Bhatt, Gen. Rel. Grav.
{\bf 7} (1976) 701.
\s
\item{[5]} J. Stachel, Phys. Rev. {\bf D 21} (1980) 2171.
\s
\item{[6]} P.S. Letetier, Phys. Rev. {\bf D 20} (1979) 1274.
\s
\item{[7]} D.R. Matraners, Gen. Rel. Grav. {\bf 20} (1988) 279.
\s
\item{[8]} J.M. Nevin, Gen. Rel. Grav. {\bf 23} (1991) 253.
\s
\item{[9]} N. Dadhich, Machian vacuum spacetime - to be published.
\s
\item{[10]} ---, Einstein Versus Newton: zero of gravitational potential
 - to be published.
\s
\item{[11]} L.K. Patel and N. Dadhich, J. Math. Phys. {\bf 34} (1993) 1927.
\s
\item{[12]} L.K. Patel and N. Dadhich, Astrophys. J. {\bf 401} (1992) 433.
\s
\item{[13]} L.K. Patel and N. Dadhich, Class. Quantum Grav. {\bf 10}
(1993) L85.
\s
\item{[14]}  R. Tikekar, L.K. Patel and N. Dadhich, Gen. Rel. Grav.
{\bf 26} (1994) 647.
\s
\item{[15]} P.C. Vaidya, Current Science {\bf 45} (1976) 490.
\s
\item{[16]} R. Adler, M. Bazin and M. Schiffer, Introduction to General
Relativity (MacGraw-Hill 1975) pp. 254-56.
\s
\item{[17]} N. Dadhich, - to be published.

\vfill\eject

\c{\bf Appendix - I}
\b
\b
$$ R_{23} = 0 $$

$$ R_{24} = (g/M) \bigg[(M_x/M)_y - (f/M^2)_u \bigg] $$

$$ R_{34} = -(g/M) \bigg[ (M_x/M)_u + (f/M^2)_y \bigg] $$

$$ R_{44} = (2/M) \bigg[M_{xx} - f^2/M^3 \bigg] $$

$$ R_{14} = (2/M) \bigg[M_{xu} + (LM_x)_x + (L f^2/M^3) \bigg] + L_{xx} $$

$$ R_{12} = LR_{24} + (g/M) \bigg[ (L_x + M_u/M)_y + (2fL/M^2)_u \bigg] $$

$$ R_{13} = LR_{34} + (g/M) \bigg[- (L_x + M_u/M)_u + (2fL/M^2)_y \bigg] $$

$$ \eqalign{R_{22} &= R_{33} = (1/M^2) \bigg[ g^2 (M_u/M)_u +
g^2 (M_y/M)_y - 1 \cr
&+ 2f (My/M) + 4(f^2L/M^2) - (M^2)_{ux} - \lbrace L(M^2)_x \rbrace_x
\bigg] \cr} $$

$$\eqalign{ R_{11} &= L^2 R_{44} + (1/M^2) \bigg[ g^2 (L_{uu} + L_{yy})
+ 2f Ly \cr
&+ 2 L_u M M _x + 4 L MM_{xu} - 2L_x MM_u + 2MM_{uu} \bigg] \cr} $$

\vfill\eject

\c{\bf Appendix II}
\b
\b
$$ \eqalign{R_{11} &= {g^2 Y \over f (X^2 + Y^2)} \bigg[ \bigtriangledown^2
G - {1 \over 2} X \bigg\lbrace ({Y_u \over Y})_{uu} + ({Y_u \over Y} )_{yy}
\bigg\rbrace \cr
&- X_u ({Y_u \over Y})_u - Y_u ({Y_u \over Y})_y \bigg] \cr
&+ {1 \over (X^2 + Y^2)} \bigg[ 3F {Y_u \over Y} - 2 X G_u + 2Y G_y - 2
F_u - 2 X^2 Y^2_u \cr
&+ X^2 ({Y_u \over Y})_u - XY ({Y_u \over Y})_y + 2XX_{uu} + 2YY_{uu}
\bigg] \cr
&+ {1 \over (X^2 + Y^2)^2} \bigg[ 2E (XY_u - YX_u) - Y^2_u (X^2 + 3Y^2)
+ 2Y^2
 X^2_u \cr
&- 2X (X^2 + 2Y^2) X_u ({Y_u \over Y}) + 2GX (X^2 - Y^2)
({Y_u \over Y}) \cr
&- 4 GY^2 X_u \bigg] - ({Y_u \over Y})_u \cr} $$
\bye